\documentclass{webofc}

\usepackage[varg]{txfonts}   
\usepackage{hyperref}
\usepackage{url}
\usepackage{graphicx} 
\usepackage{bm}
\usepackage{amssymb}
\usepackage{amsmath}
\usepackage{color}
\usepackage{cancel}
\usepackage{comment}
\usepackage{here}
\usepackage{float}
\usepackage{mathrsfs}
\usepackage{slashed}
\usepackage{mathtools}
\usepackage{physics}
\usepackage{subfigure}
\usepackage{ulem}
\usepackage{silence}
\hypersetup{colorlinks=true,citecolor=blue,urlcolor=blue,linkcolor=blue}
\begin{document}
\title{Testing the gauged $\mathrm{U(1)}_{B-L}$ model for loop induced neutrino mass with dark matter \thanks{OU-HET-1246}}
%
%

\author{\firstname{Guohao} \lastname{Ying}\inst{1} \fnsep\thanks{speaker} \and \firstname{Shinya}
\lastname{Kanemura}\inst{1} \and
        \firstname{Yushi} \lastname{Mura}\inst{1}    
}

\institute{Department of Physics, Osaka University, Toyonaka, Osaka 560-0043, Japan}

\abstract{We present a new viable benchmark scenario under the current experimental data for the model which can explain tiny mass of active neutrinos and dark matter, as a summary of Ref.~\cite{Kanemura:toappear}. Majorana masses of right-handed neutrinos are given by the spontaneous breaking of the $\mathrm{U(1)}_{B-L}$ gauge symmetry above the electroweak scale, and tiny neutrino masses are radiatively induced by quantum effects of particles of the dark sector including dark matter candidates. We first show benchmark points which satisfy current experimental data, and then give comments on how this model can be tested at collider experiments.
}
\maketitle
\section{Introduction}
\label{intro}
With the discovery of the Higgs boson at the Large Hadron Collider (LHC) in 2012, all the fundamental particles in the Standard Model (SM) of particle physics have been verified by experiments. 
However, there are some phenomena that the SM cannot explain, such as the origin of tiny neutrino masses, the nature of dark matter and the matter-antimatter asymmetry in the universe.

In order to explain the neutrino mass, Seesaw mechanism is proposed. The type-I seesaw model~\cite{Yanagida:1979as,Gell-Mann:1979vob,Minkowski:1977sc} can explain neutrino masses with very heavy right-handed neutrinos or very small Yukawa couplings. 
On the other hand, in order to explain tiny neutrino and dark matter simultaneously, radiative seesaw models have been widely studied, in which dark matter candidates are running in the loop of neutrino masses.
For example, the model proposed by Tao and Ma~\cite{Tao:1996vb,Ma:2006km} can explain observed neutrino masses with TeV scale dark matter.
However, this model cannot explain the mass origin of right-handed (RH) neutrinos, and RH neutrinos cannot become the DM candidate due to strong constraints from lepton flavor violation (LFV)~\cite{Kubo:2006yx}. 
RH neutrino DM scenario can be realized in the radiative seesaw model with gauged $\mathrm{U(1)}_{B-L}\times\mathbb{Z}_2$ extension~\cite{Kanemura:2011vm}. 
RH neutrinos receive their masses through the spontaneous symmetry breaking (SSB) of the $\mathrm{U(1)_{B-L}}$ gauge symmetry~\cite{Okada:2010wd}. 
The observed DM relic abundance can be realized through the pair annihilation via the $s$-channel scalar exchange by the mixing between the SM Higgs field and the extra scalar singlet. 
The stability of DM is guaranteed by the unbroken $\mathbb{Z}_2$ symmetry. 
Tiny neutrino masses are generated radiatively via the one-loop induced operator.

\section{The radiative seesaw model with gauged \texorpdfstring{$\mathrm{U(1)}_{B-L}$}{U(1)B-L} extension}
\label{sec-1}

The radiative seesaw model with gauged $\mathrm{U(1)}_{B-L}\times\mathbb{Z}_2$ extension~\cite{Kanemura:2011vm} is a simple extension of the SM which can radiatively generate neutrino masses and provide possible dark matter candidates. 
This model extends the SM with three $\mathbb{Z}_2$ odd right-handed neutrinos $N_\alpha$  $(\alpha=1,2,3)$, a $\mathbb{Z}_2$ odd $\mathrm{SU(2)}_L$ scalar doublet $\eta$, a scalar singlet $S$ and an electrically neutral $\mathrm{U}(1)_{B-L}$ gauge boson $Z^\prime$.
The Lagrangian is invariant under the group $\mathrm{SU(3)}_C\times\mathrm{SU(2)}_L\times\mathrm{U(1)}_Y\times\mathrm{U(1)}_{B-L}\times\mathbb{Z}_2$, where the $\mathbb{Z}_2$ symmetry is unbroken. The particle content is shown in Table~\ref{tab:tab-1}.
\begin{table}[t]
    \centering
    \begin{tabular}{c|c c c c c c|c c c}
        \hline
        {}&$Q^i$&$u_R^i$&$d_R^i$&$L^i$&$e_R^i$&$\Phi$&$N_\alpha$&$\eta$&$S$\\
        \hline
        $\mathrm{SU(3)}_C$&$\mathbf{3}$&$\mathbf{3}$&$\mathbf{3}$&$\mathbf{1}$&$\mathbf{1}$&$\mathbf{1}$&$\mathbf{1}$&$\mathbf{1}$&$\mathbf{1}$\\
        \hline
        $\mathrm{SU(2)}_L$&$\mathbf{2}$&$\mathbf{1}$&$\mathbf{1}$&$\mathbf{2}$&$\mathbf{1}$&$\mathbf{2}$&$\mathbf{1}$&$\mathbf{2}$&$\mathbf{1}$\\
        \hline
        $\mathrm{U(1)}_Y$&$\frac{1}{6}$&$\frac{2}{3}$&$-\frac{1}{3}$&$-\frac{1}{2}$&$-1$&$\frac{1}{2}$&0&$\frac{1}{2}$&0\\
        \hline
        $\mathrm{U(1)}_{B-L}$&$\frac{1}{3}$&$\frac{1}{3}$&$\frac{1}{3}$&$-1$&$-1$&0&$-1$&0&$+2$\\
        \hline
        $\mathbb{Z}_2$&+&+&+&+&+&+&$-$&$-$&+\\
        \hline
    \end{tabular}
    \caption{Particle content and their quantum numbers.}
    \label{tab:tab-1}
\end{table}

The relevant interaction Lagrangian $\mathcal{L}_\mathrm{int}$ is 
\begin{equation}
    \mathcal{L}_\mathrm{int} = \mathcal{L}^\mathrm{SM}_\mathrm{Yukawa}+\mathcal{L}_N-V(\Phi,\eta,S),
\end{equation}
where $\mathcal{L}^\mathrm{SM}_\mathrm{Yukawa}$ is the SM Yukawa interaction, and 
\begin{equation}
    \mathcal{L}_N=\sum_{\alpha=1}^3\qty(-\sum_{i=1}^3g_{i\alpha}\overline{L_i}\tilde{\eta}N_\alpha-\dfrac{y^R_\alpha}{2}\overline{N^c_\alpha}SN_\alpha+\mathrm{h.c.})
\end{equation}
with $\tilde{\eta}=i\tau^2\eta^*$, where $i,\alpha = 1,2,3$ are flavors of leptons.
Without loss of generality, the Yukawa coupling $y^R$ of RH neutrinos can be flavor diagonal. 
The Yukawa coupling among $\overline{L_i}\Phi N_\alpha$ is prohibited due to the unbroken $\mathbb{Z}_2$ symmetry. 
Masses of active neutrinos remain massless at tree level. 

Scalar fields in this model can be parameterized as 
\begin{equation}
    \Phi =
    \begin{pmatrix}
        G^+ \\
        \dfrac{1}{\sqrt{2}}\qty(v+\phi+iz)
    \end{pmatrix},\ 
    \eta = \begin{pmatrix}
        H^+\\
        \dfrac{1}{\sqrt{2}}(H+iA)
    \end{pmatrix}\ \mathrm{and}\ 
    S = \dfrac{1}{\sqrt{2}}\qty(v_S+\phi_S+iz_S).
\end{equation}
The scalar potential $V(\Phi,\eta,S)$ for this model is given by
\begin{equation}
    \begin{aligned}
        V(\Phi,\eta,S) &= \mu_1^2\abs{\Phi}^2 + \mu_2^2\abs{\eta}^2 + \mu_S^2\abs{S}^2 + \dfrac{\lambda_1}{2}\abs{\Phi}^4 + \dfrac{\lambda_2}{2}\abs{\eta}^4 + \lambda_S\abs{S}^4\\
        &+ \lambda_3\abs{\Phi}^2\abs{\eta}^2 + \lambda_4\abs{\Phi^\dagger\eta}^2 + \dfrac{\lambda_5}{2}\qty[(\Phi^\dagger\eta)^2 + \mathrm{h.c.}]+\tilde{\lambda}\abs{\Phi}^2\abs{S}^2+\lambda\abs{\eta}^2\abs{S}^2,
    \end{aligned}
    \label{eq:scalar_potential}
\end{equation}
where $\lambda_5$ can be taken as real. 
We assume $\mu_1^2 < 0,\ \mu_S^2 < 0,\ \mu_2^2>0$, that $\Phi$ and $S$ will receive non-zero vacuum expectation values (VEVs) $v$ and $v_S$ after their spontaneous symmetry breaking, respectively.
The VEV of $\eta$ remains zero. 
The mixing between $\phi$ and $\phi_S$ leads to the following mass terms
\begin{align}
\dfrac{1}{2}\begin{pmatrix}
        \phi & \phi_S
    \end{pmatrix}
    \mathcal{M}^2
    \begin{pmatrix}
        \phi\\
        \phi_S
    \end{pmatrix} =
    \dfrac{1}{2}\begin{pmatrix}
        \phi & \phi_S
    \end{pmatrix}
    \begin{pmatrix}
        \lambda_1v^2 & \widetilde{\lambda}vv_S\\
        \widetilde{\lambda}vv_S & 2\lambda_Sv_S^2
    \end{pmatrix}
    \begin{pmatrix}
        \phi\\
        \phi_S
    \end{pmatrix}.
\end{align}
The squared mass matrix $\mathcal{M}^2$ can be diagonalized by an orthogonal matrix with the mixing angle~$\alpha$
\begin{equation}
    \begin{pmatrix}
        h_1\\
        h_2
    \end{pmatrix}=
    \begin{pmatrix}
        \cos\alpha & -\sin\alpha\\
        \sin\alpha & \cos\alpha
    \end{pmatrix}
    \begin{pmatrix}
        \phi\\
        \phi_S
    \end{pmatrix},
\end{equation}
where $h_1$ and $h_2$ are mass eigenstates of Higgs bosons.

\begin{figure}[t]
    \centering
    \includegraphics[scale = 0.9,clip]{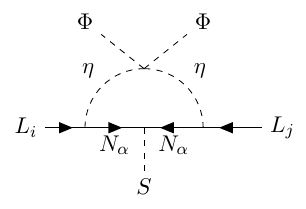}
    \caption{The one-loop diagram which generates small neutrino masses.}
    \label{fig:fig-1}
\end{figure}

In this model, tiny neutrino masses are generated via the one-loop induced dimension-six operator $LLS\Phi\Phi/\Lambda^2$, see FIG.~\ref{fig:fig-1}.
Following the framework of Ref.~\cite{Tao:1996vb, Ma:2006km}, after the electroweak symmetry breaking, the mass matrix of light neutrinos at one-loop level is 
\begin{align}
    m_\nu^{ij} = \sum_\alpha \dfrac{m_{N_\alpha}g_{i\alpha}g_{j\alpha}}{32\pi^2}\qty[\dfrac{m_H^2}{m_{N_\alpha}^2-m_H^2}\ln\qty(\dfrac{m_{N_\alpha}^2}{m_H^2})-\dfrac{m_A^2}{m_{N_\alpha}^2-m_A^2}\ln\qty(\dfrac{m_{N_\alpha}^2}{m_A^2})].
\end{align}
where $m_{N_\alpha},m_{H^\pm},m_H,m_A$ are masses of $N_\alpha,H^\pm,H,A$, respectively.

\section{Parameter spaces under current constraints}
\label{sec-2}
In this model, it is possible for $N_1$ to be dark matter with the mixing between the Higgs field $\Phi$ and the extra scalar singlet $S$, so $N_1$ can annihilate via $N_1N_1\to h_1(h_2)\to f\bar{f},ZZ,W^+W^-,h_1(h_2)h_1(h_2)$ processes~\cite{Okada:2010wd}. 
The spin independent cross section $\sigma_{\mathrm{SI}}^{p}$ of $N_1$ DM has a simple relation with BSM parameters:
\begin{equation}
    \sigma_{\mathrm{SI}}^{p}\propto\qty(\dfrac{m_{N_1}\sin(2\alpha)}{v_S})^2\times\qty(\dfrac{1}{m_{h_1}^2}-\dfrac{1}{m_{h_2}^2})^2.
\end{equation}

\begin{figure}[t]
    \centering
    \includegraphics[scale = 0.65,clip]{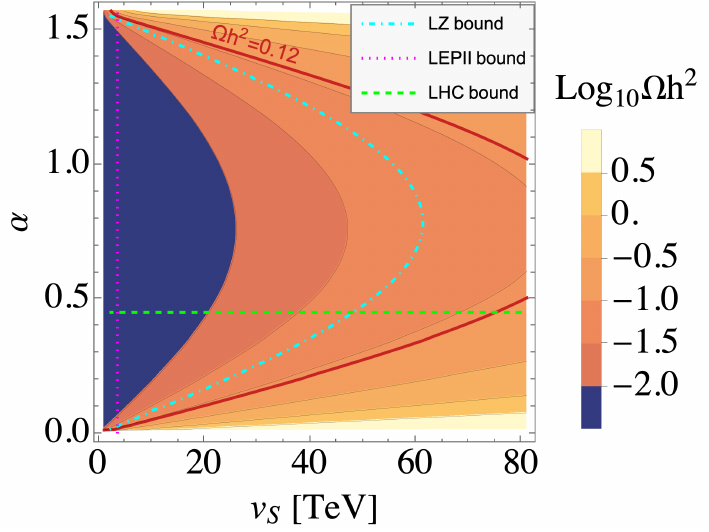}
    \caption{Parameter space of $m_{N_1}=110.4\ \mathrm{GeV}$ with constraints from LEPII~\cite{Electroweak:2003ram,Carena:2004xs}, ATLAS~\cite{ATLAS:2022vkf}, CMS~\cite{CMS:2022dwd} and LZ~\cite{LZ:2022lsv} experiments. The red line is the bound from the Planck experiment~\cite{Planck:2018vyg}.}
    \label{fig:fig-2}
\end{figure}

In FIG.~\ref{fig:fig-2}, the relic abundance of $N_1$ DM is shown as colored contour in $v_S$-$\alpha$ plane. We follow the benchmark point at
\begin{align}
    &m_{N_2} = 3500\ \mathrm{GeV},\ m_{N_3} = 4000\ \mathrm{GeV},\ \lambda = 0.01,\ \lambda_3 = 0.1,\notag \\
    &m_H = 9\ \mathrm{TeV},\ m_{H^\pm} = m_A,\ \delta\equiv m_{H^\pm} - m_H = 10^{-5}\ \mathrm{GeV}.
    \label{eq:base_bench}
\end{align}
The mass of $h_2$ is fixed at 220 GeV, and the mass of $N_1$ is 110.4~GeV near $m_{h_2}/2$. 
The allowed parameter space is the triangle encircled by the LZ 2022~\cite{LZ:2022lsv} (cyan dash-dotted), LHC~\cite{ATLAS:2022vkf,CMS:2022dwd} (green dashed) and Planck experiment~\cite{Planck:2018vyg} (red solid), which is on the right side of the LEPII constraint~\cite{Electroweak:2003ram,Carena:2004xs} (magenta dotted). 

In $N_1$ DM scenario, $N_1$ can be detected through $H^\pm$ decay processes $H^\pm\to l^\pm N_1$ in hadron and lepton colliders. 
Therefore, cross sections which produce $H^+H^-$ pairs in hadron colliders are applicable to $N_1$ DM scenario.
Inert scalar particles can be produced through monojet processes (e.g., $q\bar{q}\to Z^0g\to HAj$ and $q\bar{q}\to h_1g\to HHg$) in hadron colliders. 
Cross sections of processes including the $Z^0HA$ vertex only depend on $m_H$ and $m_A$. 
Cross section of processes including the $h_1HH$ vertex not only depend on $m_H$, but also the dimensionless parameter $\lambda_3 + \lambda_4 + \lambda_5$. 
Detailed analysis of inert scalar particles in LHC can be found in Ref.~\cite{Belyaev:2016lok}. 
Inert scalars can also be produced in lepton colliders through $e^+e^-\to Z^0\to AH(H^+H^-)$ processes~\cite{Aoki:2013lhm}. 
 
The existence of the $Z^\prime$ boson is another difference from Tao-Ma model. 
Though searches in the LHC give strong constraints on the mass of $Z^\prime$ boson, it can be lighter with smaller gauge couplings. 
Since hadron colliders (e.g., LHC and HL-LHC) have large backgrounds, it can only reach the gauge coupling $g_{B-L}$ of $\mathcal{O}(10^{-2})$. 
However, lepton colliders (e.g., ILC) can reach smaller gauge coupling of $\mathcal{O}(10^{-3})$~\cite{Basso:2009hf}. 
Decays of the $Z^\prime$ boson into SM particles are proportional to $(B-L)^2$. 
Due to its large mass, the branching ratios of $Z^\prime\to q\bar{q}, l^+l^-, \nu_L\bar{\nu}_L,N_1 N_1,h_1h_1,h_1h_2$ and $h_2h_2$ are given approximately by 0.20, 0.29, 0.15, 0.15, $6.9\times10^{-4}$, $4.4\times10^{-2}$, 0.17,  respectively~\cite{Carena:2004xs,Kanemura:2011vm}.

\section*{Acknowledgements}
The work of S.~K. was supported by the JSPS Grants-in-Aid for Scientific Research No.~20H00160, No.~23K17691 and No. 24KF0060.
The work of Y.~M. was supported by the JSPS Grant-in-Aid for JSPS Fellows No.~23KJ1460.

\end{document}